# Impact of Doping and Geometry on Breakdown Voltage of Semi-Vertical GaN-on-Si MOS Capacitors

D. Favero[a,*], C. De Santi[a], K. Mukherjee[a], M. Borga[b], K. Geens[b], U. Chatterjee[b], B. Bakeroot[c], S. Decoutere[b], F. Rampazzo[a], G. Meneghesso[a], E. Zanoni[a] and M. Meneghini[a]

[a] *Department of Information Engineering, University of Padova, Padova, Italy*
[b] *Imec, Kapeldreef 75, Heverlee 3001, Leuven, Belgium*
[c] CMST, *imec and Ghent University, Technologiepark 126, Ghent 9052, Belgium*

**Abstract**

For the development of reliable vertical GaN transistors, a detailed analysis of the robustness of the gate stack is necessary, as a function of the process parameters and material properties. To this aim, we report a detailed analysis of breakdown performance of planar GaN-on-Si MOS capacitors. The analysis is carried out on capacitors processed on different GaN bulk doping (6E18 Si/cc, 6E17 Si/cc and 2.5E18 Mg/cc, p-type), different structures (planar, trench-like) and different geometries (area, perimeter and shape). We demonstrate that (i) capacitors on p-GaN have better breakdown performance; (ii) the presence of a trench structure significantly reduces breakdown capabilities; (iii) breakdown voltage is dependent on area, with a decreasing robustness for increasing dimensions; (iv) breakdown voltage is independent of shape (rectangular, circular). TCAD simulations, in agreement with the measurements, illustrate the electric field distribution near breakdown and clarify the results obtained experimentally.

## 1. Introduction

The use of innovative semiconductor materials is a key factor to address the increasing demand for safe and reliable conversion and distribution of energy [1]. Among other materials, gallium nitride (GaN) is increasingly gaining popularity [2], thanks to the commercialization of efficient lateral transistors, with capabilities in the voltage segment up to 600-900 V [3]. However, to meet the needs of the emerging fields, such as Electric Vehicle/Hybrid Vehicles or power grids, voltage capabilities in the range of 1800 V and more are required [4]. In this context, vertical and semi-vertical GaN topologies are in the forefront of next generation power devices, overcoming the bottleneck of breakdown voltage vs area trade off affecting lateral structures [5], [6].

Among all the different architectures, trench-gate MOSFETs on foreign substrates are promising candidates for the realization of high power and high frequency switches, thanks to their economic advantages with respect to native substrates [7]. Nonetheless, heteroepitaxial vertical and semi-vertical GaN technology is in a nascent stage and several problematic issues have to be fully investigated for the realization of high-quality devices [8].

In particular, the choice of the gate stack dielectric is fundamental for the reliability of the trench MOSFET structures: a bi-layered gate dielectric has been proven to provide better performance with respect to a single-layered gate stack, without significant drawbacks [9], [10]. Other important questions still need to be addressed: how does gate stack reliability depend on the doping level and on the doping type (n-type or p-type) in the epitaxy? Does the reliability depend on gate layout geometry? What is the impact of trench formation on the reliability of the gate stack?

The aim of this paper is to address these open questions, by investigating the breakdown limits of several MOS capacitors, with different characteristics [11]. More specifically, the study is focused on the analysis of MOS capacitors with different GaN doping (n+, n and p-type), different structures (planar and trench), different shapes (rectangular and circular) and different dimensions (area and perimeter). The results, complemented by 2D simulations, provide a description of the robustness of the different layers included in the gate stack of a GaN-on-Si semi-vertical trench MOSFET.

* Corresponding author. faverodavi@dei.unipd.it

## 2. Device description

The devices under test are GaN MOS capacitors, with a bi-layered dielectric composition, grown on 8-inch silicon substrates. The study was focused on the analysis of four different wafers, with varying GaN doping and varying structures. The wafers can be classified based on their dominant processing features as follows: (i) wafer R, called "reference", planar capacitor on $n^+$ ($= 6 \cdot 10^{18}$ $cm^{-3}$) GaN; (ii) wafer L, called "lower doping", planar capacitor on n ($= 6 \cdot 10^{17}$ $cm^{-3}$) GaN; (iii) wafer P, called "p-doping", planar capacitor on p-type ($= 2.5 \cdot 10^{18}$ $cm^{-3}$) GaN; (iv) wafer T, called "trench", trench-like capacitor on $n^+$ ($= 6 \cdot 10^{18}$ $cm^{-3}$) GaN. The structures (planar and trench) and the dielectric composition ($Al_2O_3$ interface dielectric to GaN + $SiO_2$ main dielectric), are shown in Figure 1.

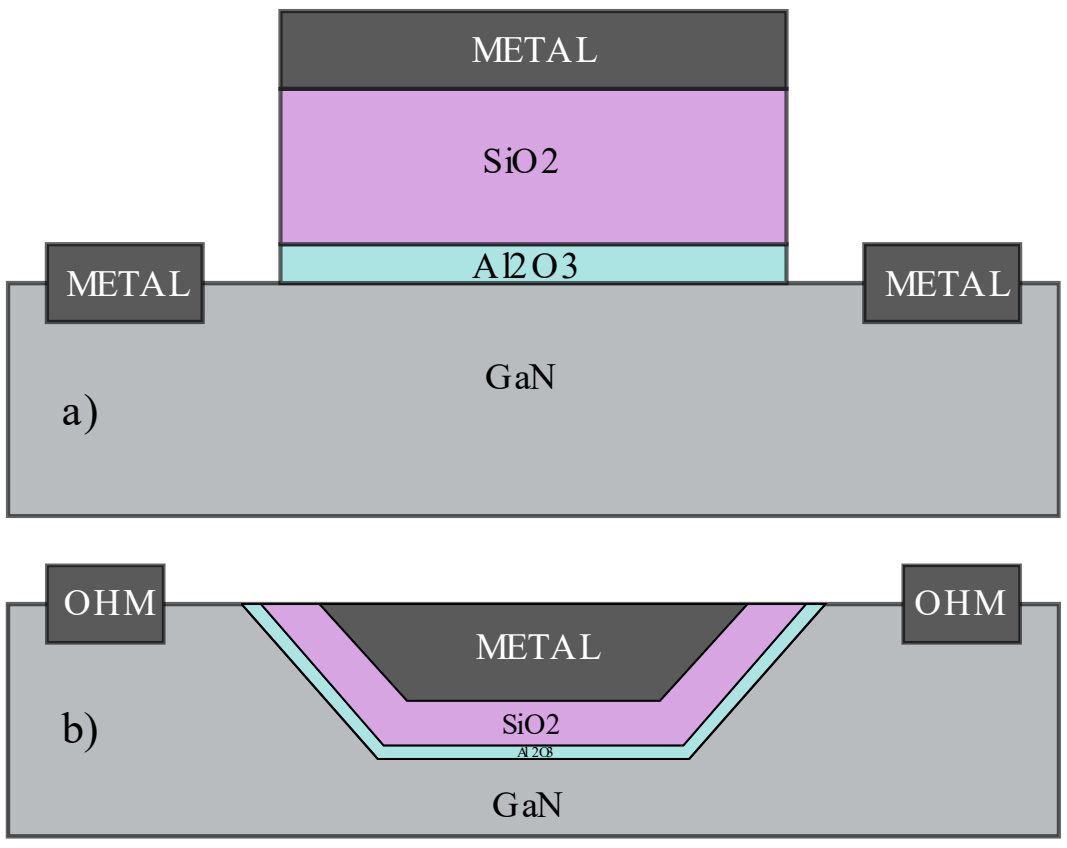

Figure 1. Schematic cross-section of the structures under test: a) Planar structure, for Wafer R, L and P; b) Trench-like structure, for wafer T. Dimensions are not to scale.

The GaN layers, in the four analysed wafers, were grown by MOCVD (metal organic chemical vapor deposition). For the wafer with a patterned trench etch, a dry etch step was used for GaN removal. First a bulk GaN removal step was implemented using a Cl2/Ar chemistry. Secondly Atomic Layer Etch (ALE) processing steps are implemented as a soft-landing step. A recess depth of ~400 nm is targeted. Before the dielectric deposition wet cleaning steps are implemented on all samples to clean up organic residues. On top of the gate dielectric a TiN/Ti/Al containing stack is used as metal electrode. Afterwards an N-implantation is used to isolate the n+-GaN in the bond pad areas of the capacitor structures. Ohmic contacts to the $n^+$-GaN are realized by recess etching, cleaning, Ti/TiN/Al based metallization and low-temperature ohmic anneal [12]. The same metallization scheme is used for the GaN stack with p-type doping, resulting in a Schottky type of contact. PECVD $SiO_2$ is used as intermetal dielectric, and the metal stacks are finished with a 4 µm-thick Al stack.

In the wafers, different capacitors were tested, based on their shape and area, to compare the breakdown performance between the different wafers and the different geometries, as discussed in the following.

In Figure 2, the typical CV characteristics of the devices are reported (voltage difference is between the metal on top of the oxide stack and the contact to the GaN). The characteristics are shown for the same circular capacitor shape, for the different analysed wafers. The figure highlights two main properties of the devices: (i) wafer T and wafer R have similar CV characteristics. This is due to the amount of doping used for gallium nitride, that is the same for the two wafers. The slight differences in the curves could be ascribed to the variability between different samples and to the different geometry. (ii) wafer L, with lower doping, has a significant lower capacitance for decreasing applied voltage. This is due to the depletion region in the semiconductor, that is wider for a lower doping device.

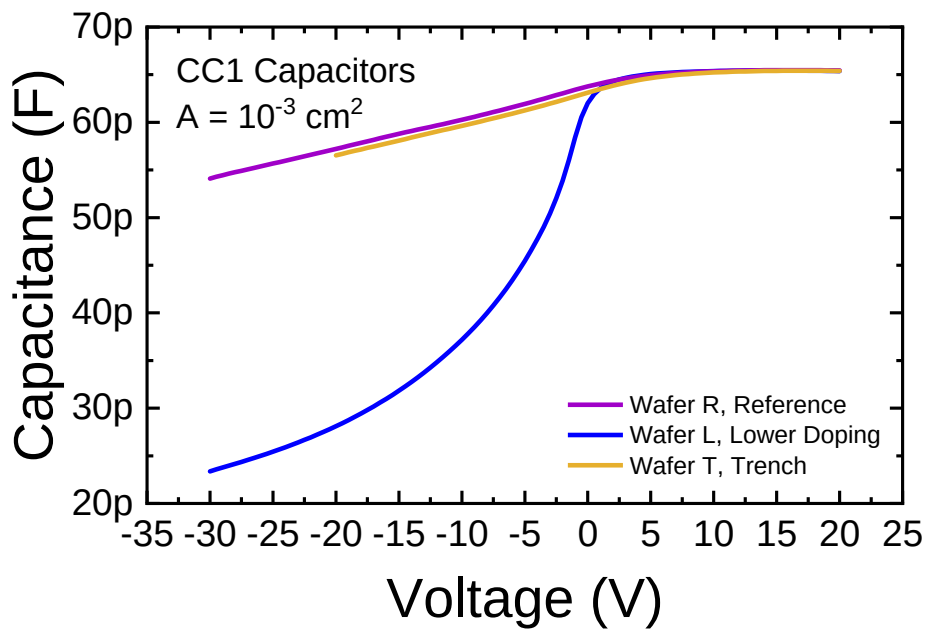

Figure 2. Typical CV characteristics of the same circular capacitor (with an area of $10^{-3}$ $cm^2$) for different wafers. Unfortunately, there was no possibility to perform CV measurements on p-doped structures, because of parasitic and non-ideal contacts on p-GaN.

For all the following tests, I-V sweeps with a semiconductor parameter analyser (Keysight E5263A) have been performed.

## 3. Breakdown dependence on doping and structure

In this section, the breakdown performance (with positive bias on the electrode on top of the oxide stack) of the MOS capacitors on the different wafers will be investigated. The analysis is carried out on identical circular capacitors with an area of $10^{-3}$ $cm^2$. The I-V characteristics until breakdown for nine identical capacitors are presented in Figure 3(a). The intra-wafer leakage level and voltage dependence are quite consistent, although the dispersion in breakdown voltage is higher in all planar capacitor wafers (see Figure 3 (b)) compared to the trench capacitors. The lowering in doping from $6 \cdot 10^{18}$ cm$^{-3}$ (wafer R) to $6 \cdot 10^{17}$ cm$^{-3}$ (wafer L) results in a worsening in breakdown voltage by 16 % (this will be discussed further). The p-doped wafer (wafer P) presents the highest average breakdown voltage (+8 % with respect to wafer R, reference). This higher breakdown voltage is ascribed to the additional voltage drop in the semiconductor, needed to bring the p-type material into inversion. The introduction of the trench etch (wafer T) reduces the breakdown voltage by 34 % (compared to wafer R). This is because of the electric field crowding at the trench corners, as shown by numerical simulations, see Figure 4.

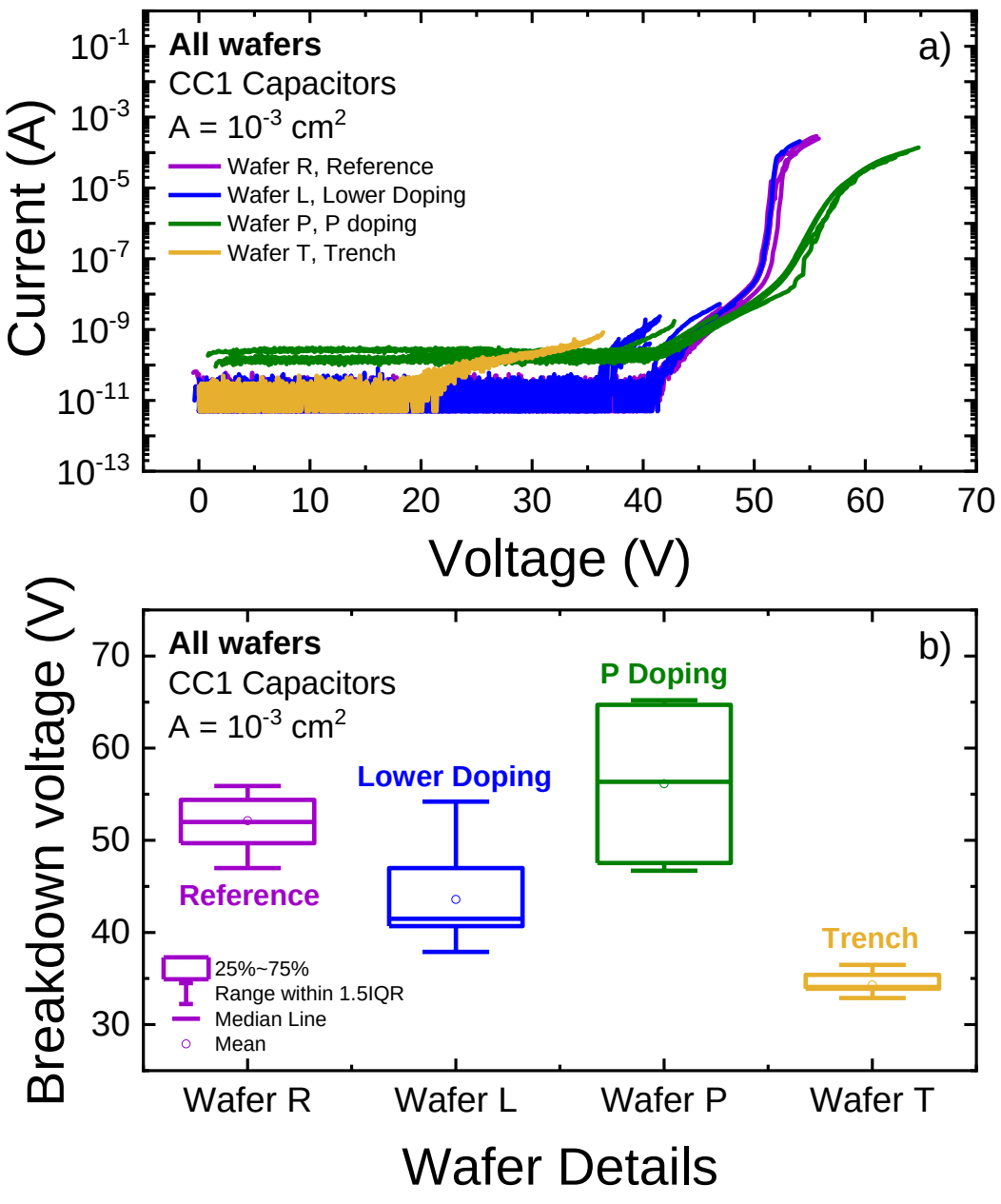

Figure 3. a) Current leakage until breakdown for the four different wafers under test, for the same circular capacitor with $10^{-3}$ cm$^2$ area; b) Breakdown voltage box plots, for all the wafers. Nine capacitors per wafer.

The breakdown results are summarized in Table 1.

| | Average VBR (V) | Minimum VBR (V) | Maximum VBR (V) | Variation respect wafer R |
|---|---|---|---|---|
| Wafer R, | 52.1 | 47 | 55.9 | - |
| Wafer L, | 43.6 | 37.9 | 54.2 | -16 % |
| Wafer P, | 56.1 | 46.7 | 65.2 | +8 % |
| Wafer T, | 34.3 | 31.5 | 36.5 | -34 % |

**Table 1.** Breakdown performance for all the wafers extracted from Figure 3.

To visualize and compare the conditions near breakdown for planar vs trench capacitor structures, Synopsys Sentaurus TCAD simulations were performed under identical bias condition (and considering ideal devices). Figure 4 presents the electric field at a representative voltage of 40 V for all planar capacitors (Figure 4 (a)) and of 35 V for trench like structures (Figure 4 (b))

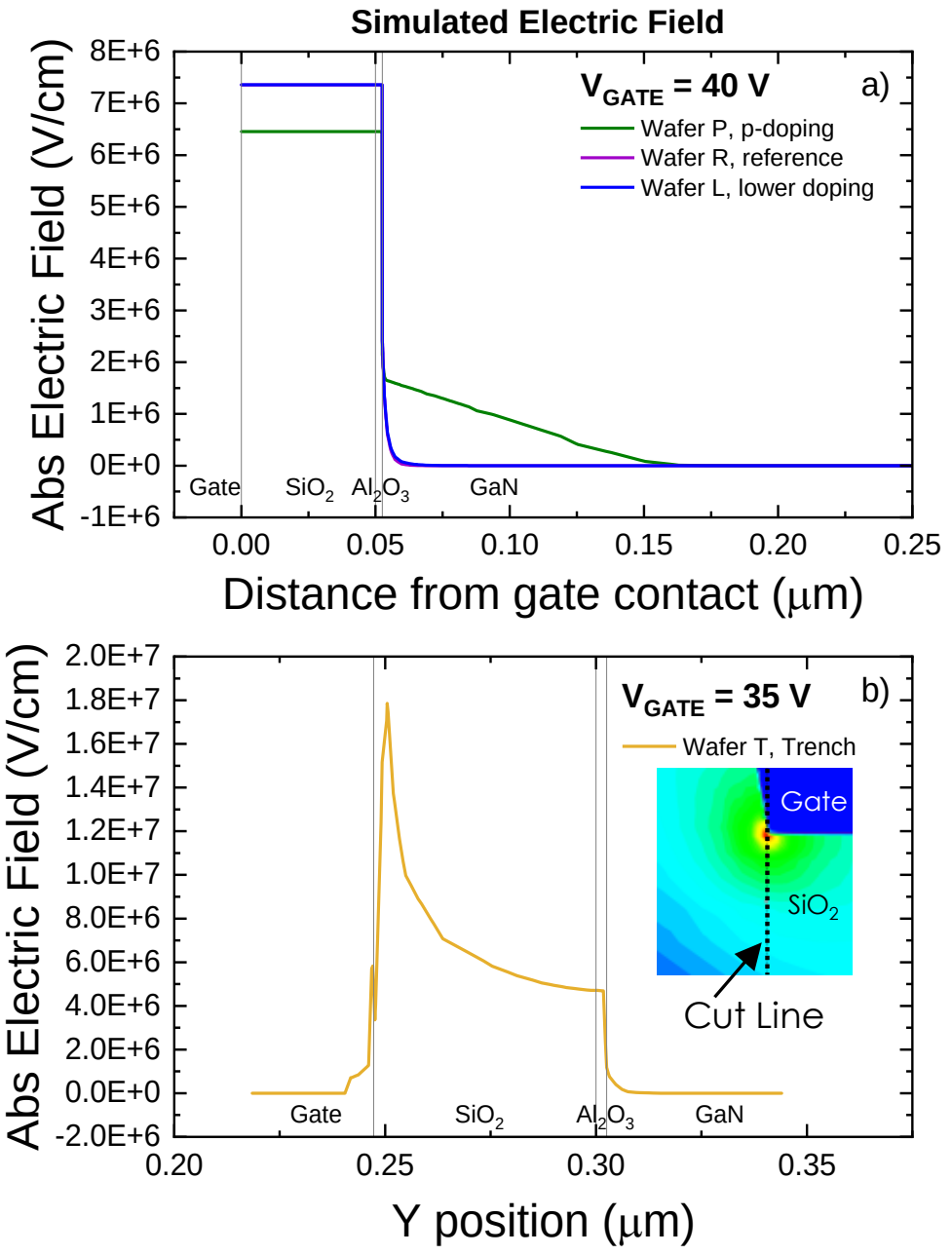

Figure 4. Simulated electric field, with two different gate voltages for planar and trench capacitors: a) planar MOS capacitors, along y direction in the middle of the gate stack, with 40 V applied (voltage close to the average breakdown of lower doping devices); b) trench MOS capacitors, along y direction under peak electric field position located close to the trench corner area (see inset), with 35 V applied (close to trench breakdown). Note that y and x scales are different in figure a) and b).

In planar capacitors the electric field distribution in the oxide is uniform across the entire gate stack, while for trench structures an evident crowding at the corner of the trench is involved, with a high and sharp electric field peak. Although these results reflect the lowering in measured breakdown voltage for trench structures, they could also explain their lower dispersion, since the failure is expected to occur in the trench corner area and thus the breakdown behaviour between samples would probably be similar. Simulations also explain the higher robustness of p-doped structures (wafer P): as previously assumed, a lower electric field is present in the oxide, due to the drop of potential in the semiconductor to get inversion. The lower electric field in the gate stack allows the p-doped structure to reach higher voltages before the occurrence of breakdown. Finally, a consideration on reference and lower-doping structures must be done: the simulated electric field is essentially the same for wafers R and L, indicating no theoretical causes (in ideal devices) for the differences observed experimentally. Although the reasons of this are not completely clear (further investigations are currently carried on), a hypothesis was formulated: the premature breakdown could be related to the higher charge trapping observed in wafer L (not shown here and currently under study), that - due to accumulation of charge in distributed defects (both in the oxide/GaN interface and in the bulk oxide) - would lead to increased electric field (with respect to reference devices) for the same applied voltage.

## 4. Breakdown dependence on geometry

The same set of I-V breakdown measurements were performed on MOS capacitors with different shape and different areas (on the same sample wafers as discussed previously), with the aim to identify any dependencies. In Figure 5 we report the breakdown voltages for circular and rectangular capacitors with two different area values: (a) devices with $A = 3 \cdot 10^{-4}$ cm$^2$; (b) devices with $A = 3 \cdot 10^{-3}$ cm$^2$.

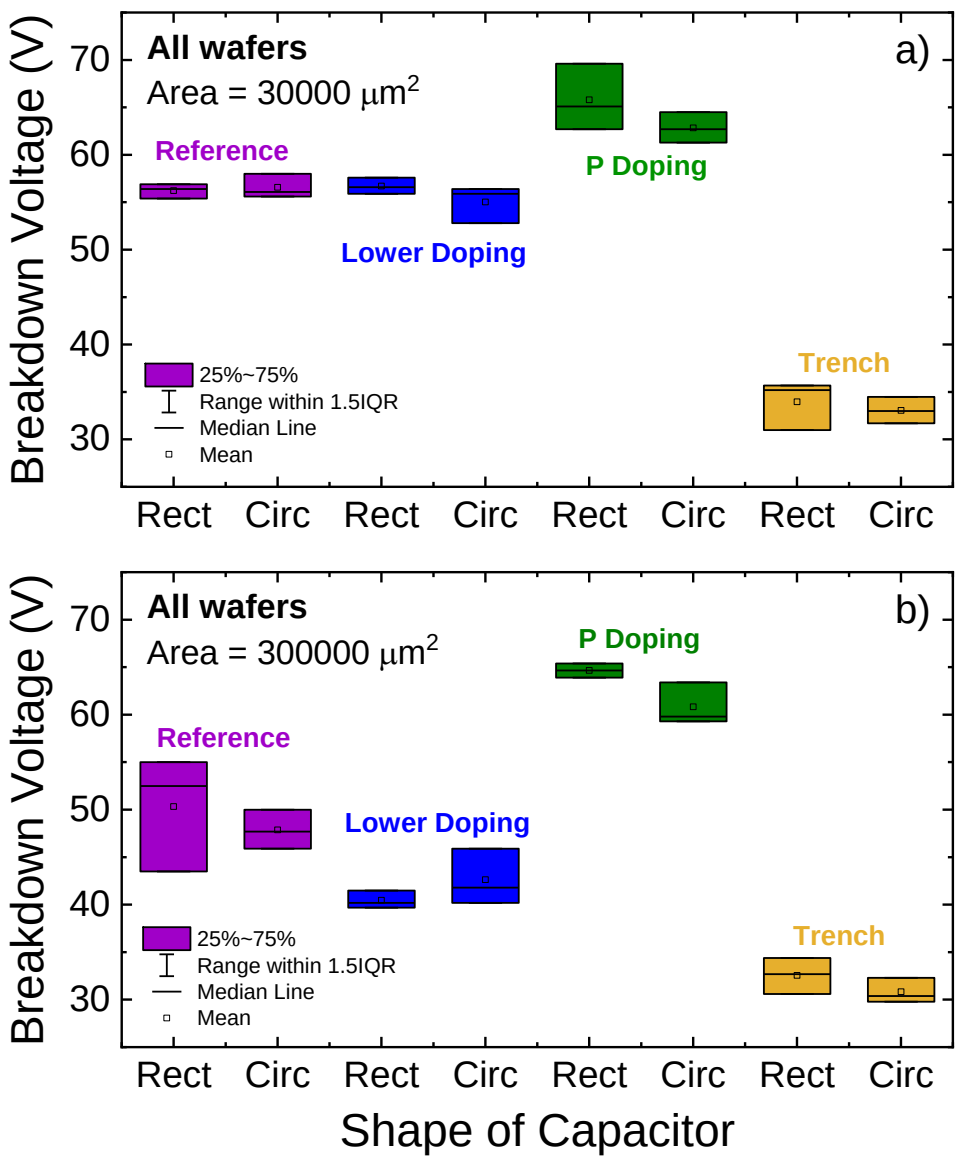

Figure 5. Breakdown dependence on shape (rectangular or circular) for two different area values: a) Devices with $\mathbf{3 \cdot 10^{-4}}$ cm$^2$ area; b) Devices with $\mathbf{3 \cdot 10^{-4}}$ cm$^2$ area. Three capacitors per box.

Capacitor shape was found not to impact the breakdown voltage. On the other hand, a clear decrease of breakdown voltage is present for increasing device area. To better analyse this phenomenon, in Figure 6 the breakdown dependence on (a) area, (b) perimeter and (c) P/A ratio, for circular capacitors, are plotted.

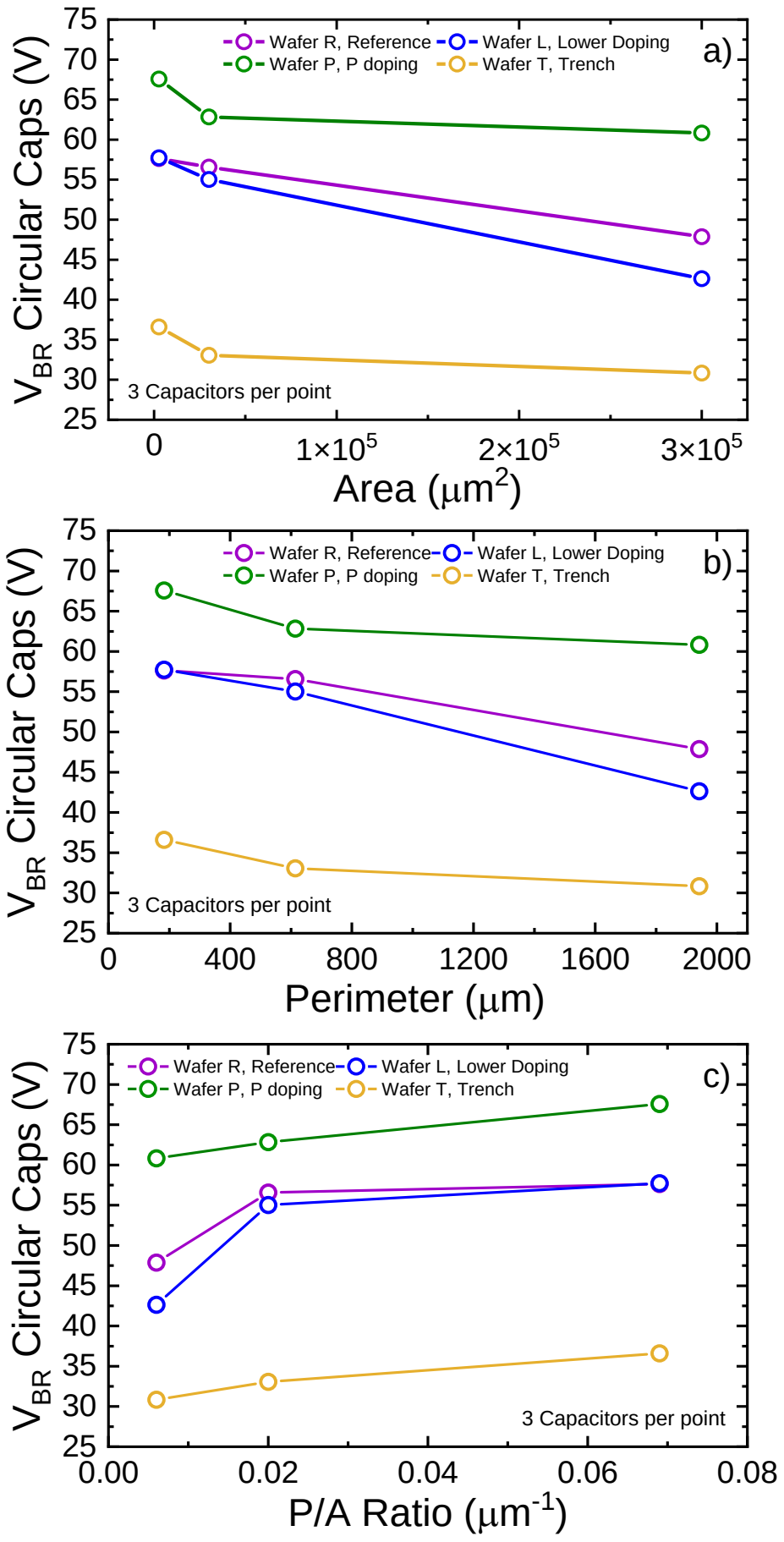

Figure 6. Breakdown dependence, for circular capacitors, on: a) Area; b) Perimeter; c) P/A ratio.

A clear trend is highlighted by this figure: for increasing area (and thus increasing perimeter since the structures considered are circular, shown in Figure 6, or rectangular, not shown) all the wafers exhibit a decreasing breakdown voltage. Considering the P/A ratio dependency, the most probable factor limiting the robustness, in this case of analysis, is the area. This could be mainly ascribed to the presence of uniformly distributed failure spots, rather than to the existence of perimeter-related defects.

## 5. Conclusions

In this paper, an extended characterization of the breakdown of GaN-on-Si MOS capacitors with a bilayer dielectric composition has been performed, as a function of the gate processing parameters. We demonstrated that: (i) material doping can significantly impact the breakdown voltage of MOS capacitors, with consequences in terms of GaN-on-Si trench MOSFETs reliability; (ii) the shape (circular, rectangular) does not play an important role for the breakdown voltage in these structures; (iii) Area is a key factor limiting the breakdown capabilities of the MOS capacitors. The results, complemented with 2D simulations, add relevant results to the research on vertical GaN MOS FET-based devices.

## Acknowledgements

This work was carried out within the UltimateGaN project, that has received funding from the ECSEL Joint Undertaking (JU) under grant agreement No 826392. The JU receives support from the European Union's Horizon 2020 research and innovation programme and Austria, Belgium, Germany, Italy, Slovakia, Spain, Sweden, Norway, Switzerland. The UltimateGaN project is co-funded by the Ministry of Education, Universities and Research in Italy

## References

[1] J. Y. Tsao *et al.*, "Ultrawide-Bandgap Semiconductors: Research Opportunities and Challenges," *Adv. Electron. Mater.*, vol. 4, no. 1, 2018, doi: 10.1002/aelm.201600501.

[2] D. Ueda, "Properties and Advantages of Gallium Nitride," in *Power GaN Devices*, Springer, 2017, pp. 1–26.

[3] B. N. Pushpakaran, A. S. Subburaj, and S. B. Bayne, "Commercial GaN-Based Power Electronic Systems: A Review," *J. Electron. Mater.*, vol. 49, no. 11, pp. 6247–6262, Nov. 2020, doi: 10.1007/s11664-020-08397-z.

[4] P. J. Wellmann, "Power Electronic Semiconductor Materials for Automotive and Energy Saving Applications - SiC, GaN, Ga 2 O 3 , and Diamond," *Zeitschrift für Anorg. und Allg. Chemie*, vol. 643, no. 21, pp. 1312–1322, Nov. 2017, doi: 10.1002/zaac.201700270.

[5] M. Meneghini *et al.*, "GaN-based power devices: Physics, reliability, and perspectives," *J. Appl. Phys.*, vol. 130, no. 18, p. 181101, Nov. 2021, doi: 10.1063/5.0061354.

[6] S. Chowdhury and D. Ji, "Vertical GaN Power Devices," in *Nitride Semiconductor Technology*, Wiley, 2020, pp. 177–197.

[7] K. Mukherjee *et al.*, "Challenges and Perspectives for Vertical GaN-on-Si Trench MOS Reliability: From Leakage Current Analysis to Gate Stack Optimization," *Materials (Basel).*, vol. 14, no. 9, p. 2316, Apr. 2021, doi: 10.3390/ma14092316.

[8] Y. Zhang, A. Dadgar, and T. Palacios, "Gallium nitride vertical power devices on foreign substrates: a review and outlook," *J. Phys. D. Appl. Phys.*, vol. 51, no. 27, p. 273001, Jul. 2018, doi: 10.1088/1361-6463/aac8aa.

[9] K. Mukherjee *et al.*, "Use of Bilayer Gate Insulator in GaN-on-Si Vertical Trench MOSFETs: Impact on Performance and Reliability," *Materials (Basel).*, vol. 13, no. 21, p. 4740, Oct. 2020, doi: 10.3390/ma13214740.

[10] D. Favero *et al.*, "Influence of Drain and Gate Potential on Gate Failure in Semi-Vertical GaN-on-Si Trench MOSFETs," *IEEE Int. Reliab. Phys. Symp.*, vol. 2, pp. 1–4, 2022, doi: 10.1109/IRPS48227.2022.9764600.

[11] M. Borga *et al.*, "Modeling of gate capacitance of GaN-based trench-gate vertical metal-oxide-semiconductor devices," *Appl. Phys. Express*, vol. 13, no. 2, p. 024006, Feb. 2020, doi: 10.35848/1882-0786/ab6ef8.

[12] B. De Jaeger, M. Van Hove, S. Decoutere, and S. Stoffels, "Low temparature ohmic contacts for III-N power devices," U.S Patent No. 9.634.107.